\documentclass[aps,prl,twocolumn,groupedaddress]{revtex4}
\usepackage{graphicx}
\def\vc#1{\mbox{\boldmath $#1$}}
\def\Be{{^{8}{\rm Be}}}
\def\C{{^{12}{\rm C}}}
\def\O{{^{16}{\rm O}}}
\def\Ne{{^{20}{\rm Ne}}}

\begin{document}


\title{Hoyle band and $\alpha$ condensation in $\C$}

\author{Y.~\textsc{Funaki}}
\email[]{funaki@riken.jp}
\affiliation{Nishina Center for Accelerator-Based Science, The Institute of Physical and Chemical Research (RIKEN), Wako 351-0198, Japan}


\date{\today}

\begin{abstract}
The excited states in $\C$ are investigated by using an extended version of the so-called Tohsaki-Horiuchi- Schuck-R\"opke (THSR) wave function, where both the $3\alpha$ condensate and $\Be + \alpha$ cluster asymptotic configurations are included. A new method is also used to resolve spurious continuum coupling with physical states.
 We focus on the structures of the ``Hoyle band'' states, $2_2^+$, and $4_2^+$ states, which are recently observed above the Hoyle state and of the $0_3^+$ and $0_4^+$ states, which are also quite recently identified in experiment. Their resonance parameters and decay properties are reasonably reproduced.
All these states have gaslike configurations of the $3\alpha$ clusters with larger root mean square radii than that of the Hoyle state. The Hoyle band is not simply considered to be the $\Be(0^+) +\alpha$ rotation as suggested by previous cluster model calculations, nor to be a rotation of a rigid-body triangle-shaped object composed of the $3\alpha$ particles. This is mainly due to the specificity of the Hoyle state, which has the $3\alpha$ condensate structure and gives a rise to the $0_3^+$ state with a prominent $\Be(0^+)+\alpha$ structure as a result of very strong monopole excitation from the Hoyle state.

\end{abstract}


\maketitle


Nuclear clustering is one of the fundamental degrees of freedom in nuclear excitation~\cite{tang}. The Hoyle state, the second $J^\pi=0^+$ state at $7.654$ MeV in $\C$, as a typical example of the cluster states, has a long history ever since it was predicted by F. Hoyle~\cite{hoyle} and subsequently observed by Cook {\it et al.}~\cite{cook} as a key state in a synthesis of $\C$ in stellar evolution. The microscopic and semi-microscopic cluster models have clarified that the Hoyle state has the structure of the $\alpha$ particle loosely coupling in an $S$-wave with the $\Be(0^+)$ core~\cite{horiuchi_ocm,kamimura,uegaki,baye}, not like a linear-chain structure of the $3\alpha$ particles proposed by Morinaga in 1950's~\cite{morinaga}. In the last decade, however, the aspect of the $\alpha$ condensate, where the $3\alpha$ clusters occupy an identical $S$-orbit, has triggered a great interest, since the so-called Tohsaki-Horiuchi-Schuck-R\"opke (THSR) wave function~\cite{thsr}, which has the $\alpha$ condensate character, was shown to be equivalent to the Hoyle state wave function obtained by solving the equations of the full $3\alpha$ resonating group method (RGM) or generator coordinate method (GCM)~\cite{funaki_2003}. 

On the other hand, the excited states of the Hoyle state have been highlighted by recent great developments in experimental studies. The second $2^+$ state $(2_2^+)$, which had been predicted at a few MeV above the Hoyle state by the cluster model calculations, was recently confirmed by many experiments~\cite{itoh_npa,itoh_prc,freer_22+,gai_first_22+,gai_22+,fynbo}. The GCM and RGM calculations propose that the $2_2^+$ state is built on the Hoyle state as a rotational member with a $\Be(0^+)+\alpha$ configuration. Freer {\it et al.} quite recently reported a new observation of the $4^+$ state at $13.3$ MeV, which they consider to compose the ``Hoyle band''~\cite{freer_42+}, together with the $0_2^+$ and $2_2^+$ states. It is proposed that this band is formed by a rotation of a rigid $3\alpha$ cluster structure with an equilateral triangle shape based on $D_{3h}$ symmetry~\cite{iachello,gai_d3h}, which is, however, not consistent with the picture of loosely coupled $\Be(0^+)+\alpha$ structure or the $3\alpha$ gaslike structure. 

Besides the $2^+$ and $4^+$ states, a $0^+$ state at $10.3$ MeV with a broad width, $\Gamma \approx 3$ MeV, has been known for a long time. However, quite recently Itoh {\it et al.} decomposed the broad $0^+$ state into the $0_3^+$ and $0_4^+$ states at $9.04$ MeV and $10.56$ MeV, with the widths of $1.45$ MeV and $1.42$ MeV, respectively~\cite{itoh_prc}. This observation of the two $0^+$ states is consistent with theoretical prediction done by using the orthogonality condition model (OCM) combined with the complex scaling method (CSM) and the analytical continuation of coupling constant (ACCC) method~\cite{kurokawa}. This was later on confirmed by another theoretical calculation using the OCM and CSM with higher numerical accuracy~\cite{ohtsubo}. 

On the other hand, in the antisymmetrized molecular dynamics (AMD)~\cite{enyo}, fermionic molecular dynamics (FMD)~\cite{neff_hoyle}, and GCM calculations~\cite{uegaki}, the observed $0_3^+$ state seems to be missing. The $0_3^+$ state given by the AMD and FMD, which may correspond to the observed $0_4^+$ state, is dominated by a linear-chain-like configuration of the $3\alpha$ clusters and is not inconsistent with the $0_3^+$ state obtained by the GCM calculation~\cite{uegaki}, or with the $0_4^+$ state in Ref.~\cite{kurokawa}, where $[\Be(2^+)\otimes l=2]_0$ configuration is dominant. It should also be mentioned that the authors in Ref.~\cite{kurokawa} claimed that the $0_3^+$ state has an $S$-wave dominant structure with more dilute density than that of the Hoyle state. These are also consistent with the observed decay properties of the $0_3^+$ and $0_4^+$ states that the former only decays into the $[\Be(0^+)\otimes l=0]_0$ channel and the latter decays into the $[\Be(2^+)\otimes l=2]_0$ channel with a sizable partial $\alpha$-decay width~\cite{itoh_jop}. 


In this Letter, we investigate the structures of the positive parity excited states above the $3\alpha$ threshold by using an extended version of the THSR wave function~\cite{zhou_ptep} so as to include $\Be+\alpha$ asymptotic configurations with a treatment of resonances. In particular, we focus on the ``Hoyle band'' (the $0_2^+$, $2_2^+$, and $4_2^+$ states\footnote{We hereafter mention the $4^+$ state at $13.3$ MeV as the $4_2^+$ state, for simplicity, though it is located lower than the $4^+$ state at $14.08$ MeV which forms the ground state rotational band.}), and the $0_3^+$ and $0_4^+$ states, together with the corresponding experimental data, though we also obtained some other positive parity excited states.

The extended version of the THSR wave function is written as follows:
\begin{eqnarray}
&&\Phi^{\rm THSR}_{JM}(\vc{B}_1,\vc{B}_2) \nonumber \\
&&={\widehat P}^J_{MK} {\cal A}\Big[ \exp \Big\{ -\sum_{i=1}^2 \mu_i \sum_{k=x,y,z} \frac{\xi_{ix}^2}{B_{ik}^2}\Big\} \phi^3(\alpha) \Big], \label{eq:1}
\end{eqnarray}
where the $\phi(\alpha)$ is an intrinsic wave function of the $\alpha$ particle, where the $(0s)^4$ configuration of the four nucleons is assumed with the size parameter $b$, which is kept fixed at $b=1.348$ fm as almost the same value as at free space. $\vc{\xi}_i$ is the Jacobi coordinates between the $3\alpha$ particles and $\mu_i=i/(i+1)$, for $i=1,2$. This is a fully microscopic wave function and every nucleons are antisymmetrized by ${\cal A}$. ${\widehat P}^J_{MK}$ is a usual angular-momentum-projection operator. This wave function is characterized by the parameters $\vc{B}_1$ and $\vc{B}_2$, which correspond to the sizes of the $\Be$ core and the remaining $\alpha$ particle center-of-mass (c.o.m.) motion, respectively. In the subsequent calculations, the axial symmetric deformation is assumed, i.e. $\vc{B}_i=(B_{ix}=B_{iy},B_{iz})$ $(i=1,2)$, for simplicity. We should note that the case of $\vc{B}_1=\vc{B}_2$ results in the original THSR wave function, where the c.o.m. motions of the $3\alpha$ particles relative to the total c.o.m. position are condensed into a lowest energy $0S$ orbit, reflecting the bosonic feature~\cite{funaki_2005,funaki_2009}. Thus this new THSR wave function is a natural extension of the original version, so that taking $|\vc{B}_1| \ll |\vc{B}_2|$ allows for the $\Be + \alpha$ cluster structure, deviating from the identical $3\alpha$-cluster motion for $\vc{B}_1=\vc{B}_2$. This new wave function still gives gaslike cluster structure, as the original THSR wave function does, not being featured by the relative distance parameter between the $\Be$ and $\alpha$ clusters. It should also be mentioned that the THSR-type wave functions were recently shown to give the best description for various cluster states such as the $\O + \alpha$ inversion doublet band in $\Ne$~\cite{zhou_prl}, $3\alpha$- and $4\alpha$-linear-chain states~\cite{suhara_prl}, and $2\alpha + \Lambda$ cluster states in ${^9_\Lambda {\rm Be}}$~\cite{funaki_9LBe}, etc.

For the excited states above the $3\alpha$ threshold, it is well known that the application of the bound state approximation gives accidental mixing between spurious continuum states and resonances. By using the fact that the root mean square (r.m.s.) radii of spurious continuum states are calculated to be extremely large within the bound state approximation, we developed a new method to remove the spurious components~\cite{funaki_PTP2005}. First we diagonalize the operator of mean square radius as follows:
\begin{eqnarray}
&&\hspace{-0.4cm} \sum_{\vc{B}_1^\prime,\vc{B}_2^\prime} \langle \Phi^{\rm THSR}_{JM}(\vc{B}_1,\vc{B}_2) |\sum_{i=1}^{12}(\vc{r}_i-\vc{X}_G)^2 |\Phi^{\rm THSR}_{JM}(\vc{B}_1^\prime,\vc{B}_2^\prime) \rangle \nonumber \\ 
&& \times g^{(\gamma)}(\vc{B}_1^\prime,\vc{B}_2^\prime)=12 \{R^{(\gamma)}\}^2 g^{(\gamma)}(\vc{B}_1,\vc{B}_2), \label{eq:cutoff1}
\end{eqnarray}
where $\vc{X}_G$ is the total c.o.m. position.
We then remove out of the present model space the eigenstates belonging to unphysically large eigenvalues. By taking the following bases,
\begin{equation}
\Phi^{(\gamma)}_{JM} = \sum_{\vc{B}_1,\vc{B}_2}g^{(\gamma)}(\vc{B}_1,\vc{B}_2) \Phi^{\rm THSR}_{JM}(\vc{B}_1,\vc{B}_2), \label{eq:cutoff2}
\end{equation}
with $\gamma$ satisfying $R^{(\gamma)} \le R_{\rm cut}$, we diagonalize Hamiltonian as follows:
\begin{equation}
\sum_{\gamma^\prime} \langle \Phi^{(\gamma)}_{JM} |H| \Phi^{(\gamma^\prime)}_{JM} \rangle f_{\lambda}^{(\gamma^\prime)} =E_\lambda f_{\lambda}^{(\gamma)}. \label{eq:eigenwf}
\end{equation}
For Hamiltonian, we adopt Volkov No.~2 force~\cite{volkov}, with the strength parameters slightly modified~\cite{ptps_68}, as effective nucleon-nucleon interaction.
The cutoff radius is now taken to be $R_{\rm cut}=6.0$ fm. 
For diagonalizing the operator of r.m.s. radius in Eq.~(\ref{eq:cutoff1}), we adopt $8^4$ mesh points for the four-parameter set, $B_{1x}=B_{1y}, B_{1z}, B_{2x}=B_{2y}, B_{2z}$, up to around $80$ fm. Since the present extended THSR wave function can include $\Be + \alpha$ asymptotic form by taking the large values of the two width parameters $\vc{B}_1$ and $\vc{B}_2$, the $\Be + \alpha$ continuum components, as well as the $3\alpha$ continuum components, can be successfully removed by imposing the cut off for the mean square radius $R^{(\gamma)} \le R_{\rm cut}$. The more details will be shown in a forthcoming full paper.

Although we could not obtain the excited states except for the $0_2^+$ and $2_2^+$ states by using the original THSR wave function~\cite{funaki_2005}, we can now obtain the other observed $0_3^+$, $0_4^+$, and $4_2^+$ states by using the present extended THSR wave function with a treatment of resonances. 
Since all these states are resonance states above the $3\alpha$ threshold, we then calculate the partial widths of the $\alpha$ particle decaying into $[\Be(I)\otimes l]_J$ channel, which we simply denote as $[I,l]_J$, based on the $R$-matrix theory~\cite{lane}, where we use the following relations,
\begin{equation}
\Gamma_{[I,l]_J}= 2P_l(ka) \gamma^2_{[I,l]_J}, \ \gamma^2_{[I,l]_J}=\frac{\hbar^2}{2\mu a} |a{\cal Y}_{[I,l]_J}(a)|^2,\label{eq:width}
\end{equation}
where $P_l(ka)$ is the penetrability calculated from the Coulomb wave functions, and $k$, $a$, and $\mu$ are the wave number of the relative motion, the channel radius, and the reduced mass, respectively. ${\cal Y}_{[I,l]_J}(r)$ is the $\alpha$ reduced width amplitude (RWA) defined below,
\begin{equation}
\hspace{-0.1cm}{\cal Y}_{[I,l]_J}(r)\hspace{-0.1cm}=\hspace{-0.1cm}\sqrt{\frac{12!}{4!8!}}\langle [\Phi_{I}(\Be),Y_{l}(\vc{\hat \xi}_2)]_{JM}\frac{\delta(\xi_2-r)}{\xi_2^2}\phi(\alpha) | \Psi^{(\lambda)}_{JM} \rangle, \label{eq:rwa}
\end{equation}
where $\Psi^{(\lambda)}_{JM}$ is the eigenfunction in Eq.~(\ref{eq:eigenwf}), $\Psi^{(\lambda)}_{JM}=\sum_\gamma f^{(\gamma)}_\lambda \Phi_{JM}^{(\gamma)}$.

\begin{table*}[htbp]
\begin{center}
\caption{The calculated binding energies measured from the $3\alpha$ threshold, $E-E_{3\alpha}^{\rm th.}$, r.m.s. radii for mass distributions, $R_{\rm rms}$, partial $\alpha$-decay widths into $\Be(0^+)$ and $\Be(2^+)$ channels, $\Gamma_{\cal.}(\Be: 0^+)=\Gamma_{[0,J]_J}$, $\Gamma_{\cal.}(\Be: 2^+)=\sum_l \Gamma_{[2,l]_J}$ in Eq.~(\ref{eq:width}), and total width, $\Gamma_{\rm cal.}({\rm total})=\Gamma_{\cal.}(\Be: 0^+)+\Be_{\cal.}(\Be: 2^+)$. The corresponding experimental data are also shown. For the $2_2^+$ state, the experimental data in Ref.~\cite{itoh_prc} and in Ref.~\cite{gai_22+} are listed at upper and lower rows, respectively. In the calculations of the $\alpha$-decay widths, proper channel radii, ranging from $4.5$ fm to $9.0$ fm depending on decay channels, are adopted. The observed energies are input in the calculations of the penetration factor $P_l(ka)$ in Eq.~(\ref{eq:width}). }\label{tab:1}
\begin{tabular}{cccccccc}
\hline\hline
\raisebox{-1.8ex}[0pt][0pt]{states} & \raisebox{-1.8ex}[0pt][0pt]{$E-E_{3\alpha}^{\rm th.}$} & \raisebox{-1.8ex}[0pt][0pt]{$R_{\rm rms}$} & \raisebox{-1.8ex}[0pt][0pt]{$\Gamma_{\rm cal.}(\Be : 0^+)$} & \raisebox{-1.8ex}[0pt][0pt]{$\Gamma_{\rm cal.}(\Be : 2^+)$} & \raisebox{-1.8ex}[0pt][0pt]{$\Gamma_{\rm cal.}({\rm total})$} & \multicolumn{2}{c}{Exp.} \\
 &  &  &  &  &  & $E-E_{3\alpha}^{\rm th.}$ & $\Gamma_{\rm exp.}$ \\
\hline
$0_2^+$ & $0.235$ & $3.73$ & $7.7\times 10^{-6}$ & $/$ & $7.7\times 10^{-7}$ & $0.3794$ & $8.5(10)\times 10^{-6}$ \\
\raisebox{-1.8ex}[0pt][0pt]{$2_2^+$} & \raisebox{-1.8ex}[0pt][0pt]{$1.62$} & \raisebox{-1.8ex}[0pt][0pt]{$3.90$} & $0.93$ & $< 10^{-6}$ & $0.93$ & \multicolumn{1}{l}{$2.57(6)$} & $1.01(15)$ \\
 &  &  & $1.1$ & $< 10^{-3}$ & $1.1$ & $2.76(11)$ & $0.800(130)$ \\
$4_2^+$ & $3.71$ & $4.51$ & $1.6$ & $0.78$ & $2.4$ & $6.0(2)$ & $1.7(2)$ \\
$0_3^+$ & $2.65$ & $4.74$ & $1.1$ & $/$ & $1.1$ & $1.77(9)$ & $1.45(18)$ \\
$0_4^+$ & $3.91$ & $4.22$ & $0.53$ & $0.04$ & $0.6$ & $3.29(6)$ & $1.42(8)$ \\
\hline
\end{tabular}
\end{center}
\end{table*}

The binding energies, which are measured from the $3\alpha$ threshold, and $\alpha$-decay widths of the five states are displayed in TABLE~\ref{tab:1} in comparison with the corresponding experimental data.
For the $0_3^+$ and $0_4^+$ states, the calculated energies are slightly higher than those of the observed $0_3^+$ and $0_4^+$ states, respectively. However, the calculated $\alpha$-decay width of the $0_3^+$ state, $\Gamma=1.1$ MeV, is in good agreement with the observed width $\Gamma=1.45$ MeV. For the $0_4^+$ state, the decay energy into the $\Be(2^+)+\alpha$ channel is very small. Therefore the partial width of this decay channel is very sensitive to the energy position. In Ref.~\cite{itoh_jop}, the authors reported a new peak at $10.8$ MeV for the $0_4^+$ state, which decays into the $\Be(2^+)+\alpha$ channel with a partial width of $0.4$ MeV. If we adopt this energy for calculating the width of the $0_4^+$ state, the partial decay width is as high as $0.12$ MeV, which is comparable to the experimental data. For the $2_2^+$ and $4_2^+$ states, the calculated energies and widths have reasonable agreement with the corresponding experimental data. In particular, the $\alpha$-decay widths of about $1$ MeV for the $2_2^+$ state and about $2$ MeV for the $4_2^+$ state are well reproduced by our calculation. 


In TABLE~\ref{tab:1}, we also show the r.m.s. radii of the five states. The r.m.s. radius of the Hoyle state, $R_{\rm rms}=3.73$ fm is the smallest, which still corresponds to very low density, i.e. $(3.73/2.4)^3=3.8$ times lower than that of the ground state. The $2_2^+$, $4_2^+$, $0_3^+$, and $0_4^+$ states have $4.3$, $6.6$, $7.7$, and $5.4$ times lower densities than that of the ground state, respectively. This means that all these states have dilute gaslike structures of the $3\alpha$ clusters, which are quite different from the rigid-body localized structure of the $3\alpha$ clusters. 

\begin{table}[htbp]
\begin{center}
\caption{The calculated $E2$ transition strengths $B(E2)$ and monopole matrix elements $M(E0)$ in units of $e^2 {\rm fm}^4$ and ${\rm fm}^2$, respectively. The observed data are also shown for the four transitions at the right column. }\label{tab:2}
\begin{tabular}{cclccc}
\hline\hline
 & ${\rm Cal.}$ &  &  & ${\rm Cal.}$ & ${\rm Exp.}$ \\
\hline
$B(E2; 2_2^+ \rightarrow 0_2^+)$ & $295$ &  & $B(E2; 2_1^+ \rightarrow 0_1^+)$ & $9.5$ & $7.6(4)$ \\
$B(E2; 4_2^+ \rightarrow 2_2^+)$ & $591$ &  & $B(E2; 2_1^+ \rightarrow 0_2^+)$ & $0.97$ & $2.6(4)$ \\
$B(E2; 2_2^+ \rightarrow 0_3^+)$ & $104$ &  & $B(E2; 2_2^+ \rightarrow 0_1^+)$ & $2.4$ & $0.73(13)$ \\
$B(E2; 2_2^+ \rightarrow 0_4^+)$ & $27.4$ &  & $M(E0; 0_2^+ \rightarrow 0_1^+)$ & $6.4$ & $5.4(2)$ \\
$M(E0; 0_2^+ \rightarrow 0_3^+)$ & $34.5$ &  &  &  &  \\
$M(E0; 0_2^+ \rightarrow 0_4^+)$ & $0.57$ &  &  &  &  \\
\hline
\end{tabular}
\end{center}
\end{table}
Next we discuss the nature of the Hoyle band. In TABLE~\ref{tab:2}, we show the $E2$ transition strengths and monopole matrix elements. We can see the very strong $E2$ transitions between the $4_2^+$ and $2_2^+$ states, and between the $2_2^+$ and $0_2^+$ states. The transitions between the $2_2^+$ and $0_3^+$ states and between the $2_2^+$ and $0_4^+$ states are three times and ten times weaker than the one between the $2_2^+$ and $0_2^+$ states, respectively. This allows us to consider the $0_2^+$, $2_2^+$, and $4_2^+$ states form a rotational band, though the $B(E2; 2_2^+ \rightarrow 0_3^+)=104$ $e^2{\rm fm}^4$ is still strong enough, so that the $0_3^+$ state may influence the band nature.  

\begin{figure}[htbp]
\begin{center}
\includegraphics[scale=0.7]{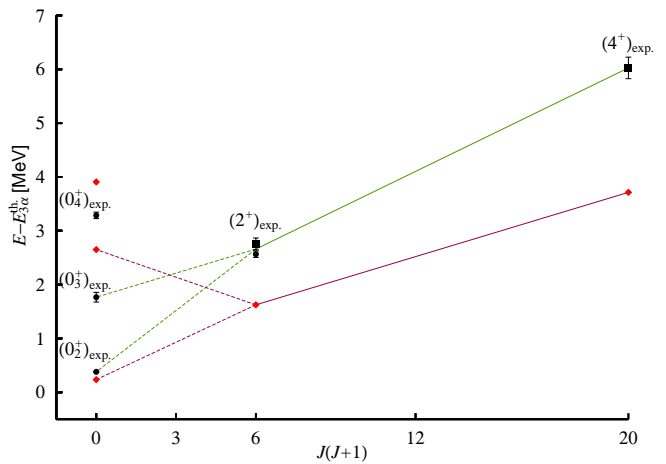}
\caption{(color online). The observed energy levels for the $0_3^+$, $0_4^+$, and $2_2^+$ states in Ref.~\cite{itoh_prc}, and the $2_2^+$~\cite{gai_22+} and $4_2^+$~\cite{freer_42+} states are denoted by black circles and black squares, respectively. The calculated energy levels for the five states are denoted by red dias.}
\label{fig:1}
\end{center}
\end{figure}

In FIG.~\ref{fig:1}, the calculated energy levels are plotted as a function of $J(J+1)$, together with the experimental data. We can say that roughly the $0_2^+$, $2_2^+$, and $4_2^+$ states both in theory and experiment follow a $J(J+1)$ trajectory. However, the $J^\pi=0^+$ bandhead in experiment seems to be fragmented into the Hoyle state and the $0_3^+$ state, and the calculated levels also have similar tendency, where the Hoyle state is located slightly below the $J(J+1)$ line. This indicates that this Hoyle band is not a simple rotational band. 

\begin{table}[htbp]
\begin{center}
\caption{The $S$-factors $S^2_{[I,l]}(J^\pi)$ defined in Eq.~(\ref{eq:sfact}) for the five $J^\pi$ states.}\label{tab:3}
\begin{tabular}{cccccccc}
\hline\hline
$J^\pi$ &  & \multicolumn{6}{c}{$S^2_{[I,l]}(J^\pi)$} \\
\hline
 \ \ & $[I,l]$\ \  & \multicolumn{2}{c}{$[0,0]$} & \multicolumn{2}{c}{$[2,2]$} & \multicolumn{2}{c}{$[4,4]$} \\
$0_2^+$\ \  &  & \multicolumn{2}{c}{$1.85$} & \multicolumn{2}{c}{$0.13$} & \multicolumn{2}{c}{$0.04$} \\
$0_3^+$\ \  &  & \multicolumn{2}{c}{$1.09$} & \multicolumn{2}{c}{$0.06$} & \multicolumn{2}{c}{$0.03$} \\
$0_4^+$\ \  &  & \multicolumn{2}{c}{$0.25$} & \multicolumn{2}{c}{$1.47$} & \multicolumn{2}{c}{$0.05$} \\
\hline
 \ \ &\ \ $[I,l]$\ \ & \ \ $[0,2]$\ \ & \ \ $[2,0]$\ \ & \ \ $[2,2]$\ \ & \ \ $[2,4]$\ \ & \ \ $[4,2]$\ \ & \ \ $[4,4]$\ \ \\
$2_2^+$\ \  & \ \  & \ \ $1.24$\ \ & \ \ $0.53$\ \ & \ \ $0.07$\ \ & \ \ $0.03$\ \ & \ \ $0.08$\ \ & \ \ $0.02$\ \ \\
\hline
 \ \ & \ \ $[I,l]$\ \ & \ \ $[0.4]$\ \ & \ \ $[4,0]$\ \ & \ \ $[2,2]$\ \ & \ \ $[2,4]$\ \ & \ \ $[4,2]$\ \ & \ \ $[4,4]$\ \ \\
$4_2^+$\ \ & \ \  & \ \ $0.88$\ \ & \ \ $0.24$\ \ &  \ \ $0.65$\ \ & \ \ $0.06$\ \ & \ \ $0.02$\ \ & \ \ $0.02$\ \ \\
\hline
\end{tabular}
\end{center}
\end{table}

We show in TABLE~\ref{tab:3} the calculated $S$-factors of the $\alpha+ \Be$ components, which can be defined below,
\begin{equation}
S_{[I,l]}(J)=\int dr \Big( r{\cal Y}_{[I,l]_J}(r) \Big)^2. \label{eq:sfact}
\end{equation}
We can see that except for the $0_4^+$ state, all the states have the largest contribution from the $[0,J]_J$ channel. This supports the idea of $\Be + \alpha$ rotation for the Hoyle band, where the $\Be$ core is in the $0^+$ ground state. 

On the other hand, the Hoyle state is considered to be the $3\alpha$ condensate state, where the $3\alpha$ clusters mutually move in an identical $S$-wave. Since the ground state of $\Be$ is composed of weakly interacting $2\alpha$ clusters coupled loosely in a relative $S$-wave, it is natural that the Hoyle state, with the $\alpha$ condensate structure, also has a large overlap with the $\Be(0^+)+\alpha$ structure. This is the same situation as for the $4\alpha$ condensate state in $\O$ discussed in Refs.~\cite{funaki_prl,enyo_16O}, which has a large overlap with the $\C(0_2^+)+\alpha$ structure. 

However, the $3\alpha$ condensate structure in the Hoyle state is not the same as the usual $\Be(0^+)+\alpha$ rotation, in which the remaining $\alpha$ cluster orbits outside the $\Be$ core. Namely in the Hoyle state, the remaining $\alpha$ cluster also orbits inside the $\Be$ core and the independent $3\alpha$-cluster motion in an identical $0S$-orbit is realized. As a result, the Hoyle state gains an extra binding, and hence its energy position is considered to be made lower than the $J(J+1)$ line, as shown in FIG.~\ref{fig:1}. The same effect is also argued in the study of the $4\alpha$ condensate and $\C(0_2^+)+\alpha$ rotational band~\cite{ohkubo,funaki_PTPS}, where the $4\alpha$ condensate is mentioned as ``complete condensate'' and the $\C(0_2^+)+\alpha$ state as ``local condensate''.  Due to the existence of the ``complete condensate'', a higher $0^+$ excited state, which has the prominent $\Be(0^+)+\alpha$ structure, with the remaining $\alpha$ cluster orbiting outside the $\Be$ core, appears as a higher nodal state, that is the $0_3^+$ state. In fact, we can see in TABLE~\ref{tab:2} that the $0_3^+$ state is strongly connected with the Hoyle state by a monopole excitation. The calculated strength $M(E0; 0_2^+ \rightarrow 0_3^+)=35$ ${\rm fm}^2$ is much larger than the other transitions, in spite of the fact that the $E0$ strength between the Hoyle and ground states $M(E0; 0_2^+ \rightarrow 0_1^+)=6.4$ ${\rm fm}^2$ is still strong enough as to be comparable to the single nucleon strength~\cite{monopole}. 
\begin{figure}[htbp]
\begin{center}
\includegraphics[scale=0.7]{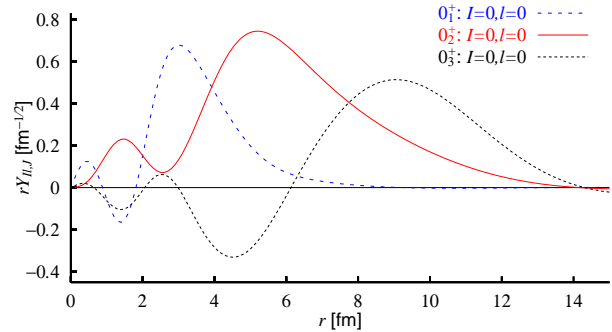}
\caption{(color online). The RWAs of the $[I,l]_J=[0,0]_0$ channel, ${\cal Y}_{[0,0]_0}(r)$ in Eq.~(\ref{eq:rwa}), for the $0_1^+$, $0_2^+$, and $0_3^+$ states.}
\label{fig:2}
\end{center}
\end{figure}

In FIG.~\ref{fig:2}, the RWAs of $[0,0]_0$ channel for the $0_2^+$ and $0_3^+$ states are shown together with that for the ground state. While the RWA for the ground state has two nodes, that for the $0_3^+$ state has four nodes and for the Hoyle state the nodal behaviour almost disappears and only a remnant of three nodes can be seen as an oscillatory behaviour. Since the outmost nodal position corresponds to a radius of repulsive core between the core $\Be$ and the $\alpha$ cluster, due to the effect of the Pauli principle, the disappearance of the nodes for the Hoyle state indicates a dissolution of the $\Be$ core, and hence formation of the $3\alpha$ condensate. On the other hand, the $0_3^+$ state, which is excited from the Hoyle state by the monopole transition, recovers the distinct nodal behaviour and, with one additional node, forms a higher nodal $\Be(0^+)+\alpha$ structure.

In TABLE~\ref{tab:3}, the $0_4^+$ state is shown to have the component of $[2,2]_0$ channel dominantly, which gives a rise to non-negligible partial decay width into this channel, consistently with the experimental information, as mentioned above. We also mention that the $2_2^+$ state also includes non-negligible mixture from the $[2,0]_2$ channel and $4_2^+$ states from the $[2,2]_4$ channel and smaller amount from the $[4,0]_4$ channel. These mixtures also deviate the $2_2^+$ and $4_2^+$ states from a pure $\Be+\alpha$ rotational structure. We will discuss this point of view in a forthcoming paper.


In conclusion, the use of the extended THSR wave function allows us to obtain the wave functions of the Hoyle band and $0_3^+$ and $0_4^+$ states, which are recently confirmed by experiments. The calculated $\alpha$-decay widths and the decay properties of these states are in good agreement with the experimental data. All these states are shown to have large r.m.s. radii and hence gaslike $3\alpha$-cluster structures. The $0_2^+$, $2_2^+$, and $4_2^+$ states are not considered to form a simple $\Be(0^+)+\alpha$ rotational band, due to the specificity of the Hoyle state with the $3\alpha$ condensate feature, which allows the $0_3^+$ state to have a prominent $\Be(0^+)+\alpha$ structure as a result of the strong monopole excitation.



\section*{Acknowledgements}
The author expresses special thanks to H. Horiuchi and A. Tohsaki for their valuable and helpful discussion and suggestions on the present work. The fruitful discussion with M. Itoh is highly appreciated. Thanks are also due to G. R\"opke, P. Schuck, T. Yamada, and B. Zhou for their stimulating discussion. This work was partially performed with the financial support by HPCI Strategic Program of Japanese MEXT, JSPS KAKENHI Grant Number 25400288, and RIKEN Incentive Research Projects.


\begin{thebibliography}{99}
\bibitem{tang}
K. Wildermuth and Y. C. Tang, {\it A Unified Theory of the Nucleus} (Vieweg, Braunschweig, 1977).
\bibitem{hoyle}
F. Hoyle, Astrophys. J. Suppl. Ser. {\bf 1}, 121 (1954).
\bibitem{cook}
C. W. Cook {\it et al.}, Phys. Rev. {\bf 107}, 508 (1957).
\bibitem{horiuchi_ocm}
H. Horiuchi, Prog. Theor. Phys. {\bf 51}, 1266 (1974); {\bf 53}, 447 (1975).
\bibitem{kamimura}
Y. Fukushima {\it et al}., Suppl. of J. Phys. Soc. Japan, {\bf 44}, 225 (1978); M. Kamimura, Nucl. Phys. A {\bf 351}, 456 (1981). 
\bibitem{uegaki}
E. Uegaki, S. Okabe, Y. Abe, and H. Tanaka, Prog. Theor. Phys. {\bf 57}, 1262 (1977); {\bf 62}, 1621 (1979).
\bibitem{baye}
P. Descouvemont and D. Baye, Phys. Rev. C {\bf 36}, 54 (1987).
\bibitem{morinaga}
H. Morinaga, Phys. Rev. {\bf 101}, 254 (1956); Phys. Lett. {\bf 21}, 78 (1966).
\bibitem{thsr}
A. Tohsaki, H. Horiuchi, P. Schuck, and G. R\"opke, Phys. Rev. Lett. {\bf 87}, 192501 (2001).
\bibitem{funaki_2003}
Y. Funaki, A. Tohsaki, H. Horiuchi, P. Schuck, and G. R\"opke, Phys. Rev. C {\bf 67}, 051306(R) (2003).
\bibitem{itoh_npa}
M. Itoh {\it et al.}, Nucl. Phys. A {\bf 738}, 268 (2004).
\bibitem{freer_22+}
M. Freer {\it et al.}, Phys. Rev. C {\bf 80}, 041303(R) (2009).
\bibitem{itoh_prc}
M. Itoh {\it et al.}, Phys. Rev. C {\bf 84}, 054308 (2011).
\bibitem{fynbo}
H.O.U. Fynbo and M. Freer, Physics {\bf 4}, 94 (2011).
\bibitem{gai_first_22+}
W.R. Zimmerman {\it et al.}, Phys. Rev. C {\bf 84}, 027304 (2011).
\bibitem{gai_22+}
W.R. Zimmerman {\it et al.}, Phys. Rev. Lett. {\bf 110}, 152502 (2013).
\bibitem{freer_42+}
M. Freer {\it et al.}, Phys. Rev. C {\bf 83}, 034314 (2011).
\bibitem{iachello}
R. Bijker and F. Iachello, Phys. Rev. C {\bf 61}, 067305 (2000); Ann. Phys. (Amsterdam) {\bf 298}, 334 (2002).
\bibitem{gai_d3h}
D. J. Mari\'n-La\'mbarri, R. Bijker, M. Freer, M. Gai, Tz. Kokalova, D.J. Parker, and C. Wheldon, Phys. Rev. Lett. {\bf 113}, 012502 (2014).
\bibitem{kurokawa}
C. Kurokawa and K. Kat${\rm \bar{o}}$, Phys. Rev. C {\bf 71}, 021301 (2005); Nucl. Phys. A {\bf 792}, 87 (2007).
\bibitem{ohtsubo}
S. Ohtsubo, Y. Fukushima, M. Kamimura, and E. Hiyama, Prog. Theor. Exp. Phys. 2013, 073D02.
\bibitem{enyo}
Y. Kanada-En'yo, Prog. Theor. Phys. {\bf 117}, 655 (2007).
\bibitem{neff_hoyle}
M. Chernykh, H. Feldmeier, T. Neff, P. von Neumann-Cosel, and A. Richter, Phys. Rev. Lett. {\bf 98}, 032501 (2007).
\bibitem{itoh_jop}
M. Itoh {\it et al.}, J. Phys.: Conf. Ser. {\bf 436}, 012006 (2013).
\bibitem{zhou_ptep}
B. Zhou, Y. Funaki, A. Tohsaki, H. Horiuchi, and Z. Z. Ren, arXiv: 1408.2920.
\bibitem{funaki_2005}
Y. Funaki, A. Tohsaki, H. Horiuchi, P. Schuck, and G. R\"{o}pke, Eur. Phys. J. A {\bf 24}, 321 (2005).
\bibitem{funaki_2009}
Y. Funaki, H. Horiuchi, W. von Oertzen, G. R\"{o}pke, P. Schuck, A. Tohsaki, and T. Yamada, 
Phys. Rev. C {\bf 80}, 064326 (2009).
\bibitem{zhou_prl}
B. Zhou, Y. Funaki, H. Horiuchi, Z. Z. Ren, G. R\"opke, P. Schuck, A. Tohsaki, C. Xu, and T. Yamada,
Phys. Rev. Lett. {\bf 110}, 262501 (2013).
\bibitem{suhara_prl}
T. Suhara, Y. Funaki, B. Zhou, H. Horiuchi, and A. Tohsaki, Phys. Rev. Lett. {\bf 112}, 062501 (2014).
\bibitem{funaki_9LBe}
Y. Funaki, T. Yamada, E. Hiyama, B. Zhou, and K. Ikeda, arXiv: 1405.6067.
\bibitem{funaki_PTP2005}
Y. Funaki, H. Horiuchi, and A. Tohsaki, Prog. Theor. Phys. {\bf 115}, 115 (2006).
\bibitem{volkov}
A. B. Volkov, Nucl. Phys. A {\bf 74}, 33 (1965). 
\bibitem{ptps_68}
Y. Fujiwara, H. Horiuchi, K. Ikeda, M. Kamimura, K. Kat${\rm \bar{o}}$, Y. Suzuki, and E. Uegaki, Suppl. Prog. Theor. Phys. {\bf 68}, 29 (1980).
\bibitem{lane}
A. M. Lane and R. G. Thomas, Rev. Mod. Phys. {\bf 30}, 257 (1958).
\bibitem{funaki_prl}
Y. Funaki, T. Yamada, H. Horiuchi, G. R\"opke, P. Schuck, and A. Tohsaki, Phys. Rev. Lett. {\bf 101}, 082502 (2008).
\bibitem{enyo_16O}
Y. Kanada-En'yo, Phys. Rev. C {\bf 89}, 024302 (2014).
\bibitem{ohkubo}
S. Ohkubo, Y. Hirabayashi, Phys. Lett. B {\bf 684}, 127 (2010).
\bibitem{funaki_PTPS}
Y. Funaki, T. Yamada, H. Horiuchi, G. R\"opke, P. Schuck, and A. Tohsaki, Suppl. Prog. Theor. Phys. {\bf 196}, 439 (2012).
\bibitem{monopole}
T. Yamada, Y. Funaki, H. Horiuchi, K. Ikeda, and A. Tohsaki, Prog. Theor. Phys. {\bf 120}, 1139 (2008).
\end{thebibliography}
\end{document}